\documentclass[11pt,a4paper]{article}
\usepackage{jcappub}
\usepackage{bm}
\usepackage{amsmath}
\def\dd{\mathrm{d}}
\def\mcH{\mathcal{H}}
\def\mcP{\mathcal{P}}
\def\mcR{\mathcal{R}}

\def\inf{{\rm inf}}

\def\Mpl{M_{\rm Pl}}
\def\GeV{{\rm GeV}}

\def\0{{(0)}}
\def\sig0{\dot{\sigma}_0}
\def\dsig{\delta \sigma}
\def\dsigd{\dot{\delta\sigma}}
\def\dN{\delta N}
\def\dphi{\delta \phi}
\def\ph0{\dot{\phi}_0}
\def\dphid{\dot{\delta \phi}}

\title{Can a spectator scalar field enhance inflationary tensor mode?}
\author[a,b]{Tomohiro Fujita}
\author[a,c]{Jun'ichi Yokoyama}
\author[d]{Shuichiro Yokoyama}
\affiliation[a]{Kavli Institute for the Physics and Mathematics of the
Universe (Kavli IPMU), WPI, TODIAS,  the University of Tokyo, 5-1-5
Kashiwanoha, Kashiwa, 277-8583, Japan}
\affiliation[b]{Department of Physics, Graduate School of Science,
The University of Tokyo, Bunkyo-ku 113-0033, Japan}
\affiliation[c]{Research Center for the Early Universe (RESCEU), Graduate School of Science, The University of Tokyo, Tokyo 113-0033, Japan}
\affiliation[d]{Department of Physics, Rikkyo University, 3-34-1 Nishi-Ikebukuro, Toshima, Tokyo, 223-8521, Japan}
\emailAdd{tomohiro.fujita@ipmu.jp}
\emailAdd{yokoyama@resceu.s.u-tokyo.ac.jp}
\emailAdd{shuichiro@rikkyo.ac.jp}

\abstract{
We consider the possibility of enhancing the inflationary tensor mode by introducing a spectator scalar field with a small sound speed
which induces gravitational waves as a second order effect. We analytically obtain the power spectra of gravitational waves and curvature perturbation induced by the spectator scalar field. 
We found that the small sound speed amplifies the curvature perturbation
much more than the tensor mode and the current observational constraint forces the induced gravitational waves to be negligible compared with those from
the vacuum fluctuation during inflation. }


\begin{document}

\maketitle

%
%
%
\section{Introduction}

Recently, cosmological primordial gravitational waves (GWs) attract much attention.
In slow-roll inflation~\cite{Starobinsky:1980te, Sato:1980yn, Guth:1980zm}
(for the latest pedagogical review of inflation, see ref.~\cite{Yokoyama:2014nga}), it is known that the stochastic gravitational waves
are generated from the vacuum fluctuation, and the power spectrum is 
proportional to the energy scale of inflation $\rho_{\rm inf}$,
\begin{equation}
\mcP_h^{\rm vac}= \frac{2 H^2}{\pi^2 \Mpl^2}
\propto \rho_{\rm inf},
\label{vac}
\end{equation}
where $H$ is the Hubble parameter during inflation and $\Mpl$ is the reduced Planck mass~\cite{Starobinsky:1979ty}.
Therefore if one measures $\mcP_h^{\rm vac}$, $\rho_\inf$ can be observationally revealed.

However, it should be noted that observing the amplitude of primordial GWs does not necessarily mean that we can determine $\rho_\inf$ immediately.
It is because not only $\mcP_h^{\rm vac}$ but also GWs from other sources possibly contribute the observed GWs. Any theoretical argument or observational
evidence guarantees that the observed GWs are dominated by $\mcP_h^{\rm vac}$.
Thus it is important to explore an alternative possibility of the generation
of GWs in the primordial universe.

In general, GWs are sourced by the anisotropic component of stress energy tensor. Since a vector field naturally provides such an anisotropic stress,
several mechanisms in which vector fields produced during inflation
source GWs are investigated~\cite{Barnaby:2010vf, Sorbo:2011rz, Barnaby:2011vw, Cook:2011hg, Mukohyama:2014gba, Ferreira:2014zia}.%
\footnote{For other mechanisms, see also ref.~\cite{Senatore:2011sp}.}
Nevertheless, the authors in ref.~\cite{Biagetti:2013kwa} pointed out
that the spatial kinetic energy of a scalar field also sources GWs
and it can be amplified when the sound speed of the scalar field
is significantly smaller than unity. 

In this paper, we consider a spectator scalar field with a generalized Lagrangian which gives a nontrivial sound speed. The GWs and the
curvature perturbation induced by the scalar field are analytically calculated.
We have found that the induced curvature perturbation becomes much larger
than the induced GWs. The requirement that the induced curvature perturbation cannot exceed the observed value puts a constraint on the amplitude of the induced GWs. Consequently, it is shown that the GWs induced by a spectator field with a small sound speed is restricted to be much smaller than $\mcP_h^{\rm vac}$.
Finally, we extend the action of the additional scalar into a more generic
form which contains the Galileon-like term. Even in this case, however, 
we found the induced GWs is strictly limited and cannot be dominant. 

The rest of the paper is organized as follows.
In Sec.~\ref{Perturbed Action}, we perturb the action and obtain the power
spectrum of the spectator field perturbation. In Sec.~\ref{Induced curvature and graviton perturbations}, the power spectra of the induced GWs and the curvature perturbation are derived and their constraints are discussed.
In Sec.~\ref{interpretation}, we develop the understanding of the reason
why such a stringent constraint on the induced GWs is obtained.
In Sec.~\ref{Galileon}, the extended action of the spectator field is briefly
argued. We conclude in Sec.~\ref{Conclusion}.

\section{Perturbed Action}
\label{Perturbed Action}

We consider the following Lagrangian:
\begin{equation}
\mathcal{L} =\frac{1}{2}\Mpl^2 R+ \frac{1}{2}\partial_\mu \phi \partial^\mu \phi
-V(\phi) + P(X, \sigma),
\label{matter lagrangian}
\end{equation}
where $\phi$ is the inflaton, $V(\phi)$ is its potential, $\sigma$ is a spectator field, and   $X \equiv \frac{1}{2}\partial_\mu \sigma \partial^\mu \sigma$. 
In this paper, the inflaton $\phi$ is assumed to be responsible for both the occurrence of inflation and the generation
of the scalar perturbations imprinted
 in the cosmic microwave background radiation 
(CMB)~\cite{Ade:2013uln, Ade:2013ydc}. 
On the other hand, the $\sigma$ field is supposed to generate gravitational waves
through its second order perturbations.
For the moment, we assume that the Lagrangian of $\sigma$ is an arbitrary
function of $X$ and $\sigma$,  $P(X,\sigma)$, while we
further extend it in Sec.~\ref{Galileon}.
In this section, we perturb the action and derive the equations of motion for the perturbed fields.

In the (3+1) decomposition, the metric is given by
\begin{equation}
\dd s^2 = N^2\dd t^2 - \gamma_{ij} (\dd x^i +N^i \dd t)(\dd x^j +N^j \dd t),
\end{equation}
where we incorporate metric perturbations around the flat FLRW metric as,
\begin{equation}
N= 1+\delta N,\quad N_i = \partial_i \psi,
\quad \gamma_{ij}= a^2 \left(\delta_{ij}+h_{ij}+\frac{1}{2}h_{ik}h^k_j\right),
\end{equation}
working in the flat gauge for the scalar perturbations
and the transverse-traceless (T.T.) gauge for the tensor perturbations.
One can show that the gravity sector of the perturbed action up to the second order  is given by
\begin{multline}
S_g^{(1,2)} = \int \dd t\dd^3 x\, a^3 \Big[\
3\Mpl^2 H^2 \delta N \quad ({\rm 1st\ order})\\
- 3\Mpl^2 H^2 (\delta N)^2 -2\Mpl^2 H\delta N a^{-2} \partial_i^2 \psi 
+\frac{\Mpl^2}{8}\left(\dot{h}_{ij}\dot{h}_{ij}-a^{-2}\partial_k h_{ij} \partial_k h_{ij}\right) \ \ {\rm (2nd \ order)}
\Big] .
 \label{gravity sector}
\end{multline}
Here we ignore the third and higher order terms in the gravity sector.
Although they include $\mathcal{O}(h \dsig^2)$ coupling terms,
these terms are slow-roll suppressed compared to similar terms in the matter sector and hence they are sub-leading~\cite{Maldacena:2002vr}.

We now consider the matter sector of the action.
The two scalar fields are decomposed into the background part and 
the perturbation, 
\begin{equation}
\phi(t,\bm{x})=\phi_0(t)+\delta\phi(t,\bm{x}),
\quad
\sigma(t,\bm{x})=\sigma_0 (t) + \delta \sigma (t,\bm{x}).
\end{equation}
The calculation of the perturbed matter action is straightforward.
First, let us compute the perturbed Lagrangian of the $\sigma$ field.
The perturbation expansion of $X \equiv \frac{1}{2}\partial_\mu \sigma \partial^\mu \sigma$ is given by
\begin{align}
&X 
= \frac{1}{2} \dot{\sigma}_0^2 \quad ({\rm 0th \ order})
\ + \sig0 \dsigd -\sig0^2 \dN \quad ({\rm 1st \ order})\notag\\
&\quad+ \frac{1}{2}\dsigd^2 -\sig0 a^{-2} \partial_i \psi \partial_i \dsig
-2\dN \sig0 \dsigd +\frac{3}{2} \sig0^2 \dN^2 -\frac{1}{2}a^{-2} (\partial_i \dsig)^2 \quad({\rm 2nd \ order})\notag\\
&\quad -\dsigd a^{-2} \partial_i \psi \partial_i \dsig -\dN \left( \dsigd^2-2\sig0 a^{-2}\partial_i \psi \partial_i \delta\sigma \right) + 3\sig0 \dsigd \dN^2\notag\\
&\quad -2\sig0^2 \dN^3 +\frac{1}{2} a^{-2} h_{ij} \partial_i \dsig\partial_j \dsig \quad ({\rm 3rd\ order}) \quad+ \mathcal{O}(\dsig^4). 
\label{X expansion}
\end{align}
%
Then one finds
\begin{align}
NP(X, \sigma)&=(1+\delta N)P(X, \sigma) \notag\\
&= P^\0 \quad ({\rm 0th\ order}) \notag\\
&\quad+ P^\0 \dN + P_X^\0 \left(\sig0 \dsigd-\sig0^2 \dN \right) + P_\sigma^\0 \dsig \quad ({\rm 1st\ order}) \notag\\
&\quad +\frac{1}{2}P_X^\0 \left[ \dsigd^2 -2\sig0 a^{-2} \partial_i \psi \partial_i \dsig -2\sig0\dsigd \dN +\sig0^2 (\dN)^2 -a^{-2}(\partial_i \dsig)^2\right]\notag\\
&\quad +\frac{1}{2} P_{XX}^\0 \left( \sig0\dsigd-\sig0^2\dN\right)^2 +\frac{1}{2}P_{\sigma\sigma}^\0
(\dsig)^2 + P_\sigma^\0 \dsig\dN \quad ({\rm 2nd\ order})\notag\\
&\quad +\frac{1}{2}P_{XX}^\0 \sig0(\dsigd)^3 - \left(\frac{1}{2} P_X^\0 -2P_{XX}^\0\sig0^2\right)(\dsigd)^2\dN
+\left(P_X^\0 +\frac{5}{2}P_{XX}^\0\sig0^2\right)\sig0\dsigd\dN^2\notag\\
&\quad - \left(\frac{1}{2}P_X^\0 +P_{XX}^\0\sig0^2\right)\sig0^2 (\dN)^3 
 + \left(P_X^\0 +P_{XX}^\0\sig0^2\right)\left(\sig0\dN-\dsigd\right) a^{-2}\partial_i\psi\partial_i \dsig
 \notag\\
&\quad -\frac{1}{2} P_{XX}^\0 \sig0\dsigd a^{-2}(\partial_i \dsig)^2 
-\frac{1}{2} \left(P_{X}^\0-P_{XX}^\0\sig0^2 \right)\dN a^{-2}(\partial_i \dsig)^2 \notag\\
&\quad +\frac{1}{2}P^{(0)}_{\sigma\sigma}(\dsig)^2\dN+\frac{1}{2} P_X^\0 h_{ij} a^{-2} \partial_i \dsig \partial_j \dsig
\quad ({\rm 3rd\ order}) \ + \mathcal{O}(\dsig^4),
\label{sigma 3rd}
\end{align}
where $P_{X^n}^\0 \equiv \partial^n P/\partial X^n |_{X=\dot{\sigma}_0^2/2, \sigma=\sigma_0}$.
Note that we suppress the terms in proportion to $P_{X\sigma}^\0, P_{\sigma\sigma\sigma}^\0$ and other higher derivatives which do not yield the $h\dsig^2$ coupling.
A general multi-field perturbed action can be found in ref.~\cite{Langlois:2008qf}
while it does not include the tensor perturbations.
One can easily obtain the perturbed lagrangian of the $\phi$ sector by
making replacements, $\dsig\to\dphi,  P_X^\0 \to 1, P_{XX}^\0\to 0, P^{(0)}_\sigma \to -V_\phi^{(0)}$
and $P^\0 \to \ph0^2/2 + V^{(0)}$ in eq.~\eqref{sigma 3rd}.

\if
$P(X, \sigma)$ is expanded as
\begin{equation}
P(X, \sigma) = P^\0+ P_X^\0 \delta X +\frac{1}{2} P_{XX}^\0 \delta X^2  +P_\sigma^\0 \delta\sigma +\frac{1}{2} P_{\sigma\sigma}^\0 \delta\sigma^2 +\cdots,
\label{P expansion}
\end{equation}
where $P_{X^n}^\0 \equiv \partial^n P/\partial X^n |_{X=\dot{\sigma}_0^2/2, \sigma=\sigma_0}$ and $\delta X\equiv X-\dot{\sigma}_0^2/2$.
The full expression of the perturbed matter lagrangian up to the third order is written in Appendix.~\ref{Full expression of the perturbed action}.
\fi

Now we have the perturbed action with the four scalar perturbation quantities,
$\dN,  \psi, \dphi$ and $\dsig$. However, the Hamiltonian and momentum constraints of the second order action eliminates the two of them,
\begin{align}
2\Mpl^2 H \dN &= \ph0\dphi + P_X^\0 \sig0 \dsig,\label{dN solution}\\
-2\Mpl^2 H a^{-2}\partial_i^2 \psi &=
\left( 6\Mpl^2 H^2 -\ph0^2-K\sig0^2\right)\dN 
+\ph0\dphid+V_\phi^{(0)}\dphi+K\sig0\dsigd-P_\sigma^\0 \dsig,
\label{psi solution}
\end{align}
with $K\equiv P_X + P_{XX}\sig0^2$.
Using these constraint equations and eliminating $\dN$ and $\psi$, we obtain the second order action
of $\dphi$ and $\dsig$ as~\cite{Langlois:2008qf}
\begin{align}
S^{(2)}= \frac{1}{2}\int \dd t\dd^3 x a^3 & \Big[ 
(\dphid)^2+K(\dsigd)^2-a^{-2}(\partial_i \dphi)^2-P_X a^{-2} (\partial_i \dsig)^2\notag\\
& -\mu_\phi^2 (\dphi)^2-\mu_\sigma^2(\dsig)^2-\Omega\dphi\dsig-\tilde{\Omega}\dphi\dsigd
\Big],
\label{2nd order phi sigma action}
\end{align}
where 
\begin{align}
\mu_\phi^2&\equiv V_{\phi\phi}+\frac{3\ph0^2}{2\Mpl^2}+\frac{\ph0 V_{\phi}}{\Mpl^2 H}
-\frac{\ph0^2}{4\Mpl^4 H^2}\left(\ph0^2+K\sig0^2\right)-\frac{\partial_t(a^3\ph0^2/H)}{2\Mpl^2a^3},\\
\mu_\sigma^2&\equiv -P_{\sigma\sigma}+\frac{3P_X^2 \sig0^2}{\Mpl^2}-\frac{P_\sigma P_X \sig0}{\Mpl^2 H} -\frac{P_X^2\sig0^2}{4\Mpl^4 H^2}\left(\ph0^2+K\sig0^2\right)-\frac{\partial_t(a^3 KP_X \sig0^2 /H)}{2\Mpl^2 a^3},\\
\Omega &\equiv 3\frac{\ph0P_X\sig0}{\Mpl^2}-\frac{P_\sigma\ph0}{\Mpl^2 H}+\frac{V_\phi P_X\sig0}{\Mpl^2 H}-\frac{\ph0P_X \sig0}{2\Mpl^4 H^2}\left(\ph0^2 +K\sig0^2\right)- \frac{\partial_t (a^3 \ph0 P_X \sig0/H)}{\Mpl^2 a^3},\\
\tilde{\Omega} &\equiv \frac{\ph0 K\sig0}{\Mpl^2 H}+ \frac{\ph0P_{XX}\sig0^3}{\Mpl^2 H}.
\end{align}
Note we omit the superscript $``(0)"$ hereafter.
To canonically normalize the fields, we redefine
\begin{equation}
\chi \equiv a\dphi,\qquad \Sigma\equiv a\sqrt{K}\dsig.
\end{equation}
With these new variables, the second-order action 
reads
\begin{align}
S^{(2)}=& \frac{1}{2}\int \dd \eta\dd^3 x   \Big[ \ 
\chi'^2-(\partial_i \chi)^2 +\left(\frac{a''}{a}-a^2\mu_\phi^2\right)\chi^2 \notag\\ &
+\Sigma'^2  - c_s^2 (\partial_i \Sigma)^2 +\left(\frac{(a\sqrt{K})''}{a\sqrt{K}}-a^2\mu_\sigma^2\right)\Sigma^2
\notag\\ &
+\frac{a}{\sqrt{K}}\left(\tilde{\Omega}\frac{(a\sqrt{K})'}{a\sqrt{K}}- a\Omega\right)\chi\Sigma
-\frac{a}{\sqrt{K}}\tilde{\Omega}\chi\Sigma'\ 
\Big],
\label{second order canonical action}
\end{align}
where the prime denotes the derivative with respect to the conformal time $\eta$ and we introduce the sound speed of the canonical field $\Sigma$,
\begin{equation}
c_s^2 \equiv \frac{P_X}{K} = \frac{P_X}{P_X+P_{XX}\sig0^2}.
\end{equation}
The equations of motion of the two canonical fields are given by
\begin{align}
\chi''-\partial_i^2\chi + \left(a^2\mu_\phi^2-\frac{a''}{a} \right)\chi
&=\frac{a}{\sqrt{K}}\left[\left(\tilde{\Omega}\frac{(a\sqrt{K})'}{a\sqrt{K}}- a\Omega\right)\Sigma
-\tilde{\Omega}\Sigma'\right],\\
\Sigma''-c_s^2 \partial_i^2 \Sigma+\left[a^2 \mu_\sigma^2-\frac{(a\sqrt{K})''}{a\sqrt{K}} \right]\Sigma &= \frac{a}{\sqrt{K}}\left(\tilde{\Omega}\frac{(a\sqrt{K})'}{a\sqrt{K}}- a\Omega\right)\chi
+ \left(\frac{a}{\sqrt{K}}\tilde{\Omega}\chi\right)'\,.
\end{align}
Since these equations are coupled to each other due to the mixing terms (see the third line in eq.~\eqref{second order canonical action}), it is hard to solve them if the mixing is significantly strong. Moreover, if the masses, $\mu_\phi^2$
and $\mu_\sigma^2$, are not much less than $H^2$, their fluctuations are not generated during inflation. Thus we explore the condition in which
both the mixing and their mass are negligible and we focus on these cases
in the following section.

The inflaton mass, $\mu_\phi^2$, is evaluated as
\begin{equation}
\frac{\mu_\phi^2}{H^2} \simeq 3\eta_\phi -6\epsilon_H+6\sqrt{\epsilon_\phi \epsilon_H}-P_X\frac{\epsilon_H^2 \sig0^2}{c_s^2\dot{\phi}^2_0} +\mathcal{O}(\epsilon^2),
\label{mu phi evaluate}
\end{equation}
where $\epsilon_H\equiv -\dot{H}/H^2,\ \epsilon_\phi\equiv \Mpl^2V_\phi^2/2V^2,\ \eta_\phi\equiv\Mpl^2V_{\phi\phi}/V$ as usual.
We also use the background equation,
$-2\Mpl^2 \dot{H} = \ph0^2 + P_X \sig0^2$.
In eq.~\eqref{mu phi evaluate}, only the last term can be large for a very small $c_s$. It requires a condition, 
\begin{equation}
c_s^2  \gg  \left|\epsilon_H^2 \frac{P_X\sig0^2}{\ph0^2}\right|,
\label{asump1}
\end{equation}
for the inflaton mass to be negligibly small. 
Provided $P_{\sigma\sigma}\lesssim V_{\phi\phi}, P_\sigma \lesssim V_\phi$
and $P_X \sig0 \lesssim \ph0$ which are natural conditions for a spectator field, one can show that eq.~\eqref{asump1} 
guarantees $\mu_\sigma^2 \ll H^2$ and $\Omega\ll H^2$.
However, for a small $c_s$, one finds
\begin{equation}
\frac{\tilde{\Omega}}{H} \simeq 4 \frac{\epsilon_H}{c_s^2} \frac{P_X \sig0}{\ph0}.
\end{equation}
To ignore the mixing, we need an additional condition;
\begin{equation}
c_s^2 \gg \left|\epsilon_H \frac{P_X \sig0}{\ph0}\right|.
\label{asump2}
\end{equation}
%
 and they do not appear if $\sigma$ has a usual kinetic term.

When the two conditions, eqs.~\eqref{asump1} and \eqref{asump2},
are satisfied and the slow-roll parameters are sufficiently small,
the mass terms and the mixing terms are safely ignored.
Then we obtain the mode functions of the two fields as
\begin{align}
\chi_k \simeq \frac{e^{-ik\eta}}{\sqrt{2k}}\left(1-\frac{i}{k\eta}\right),
\qquad
\Sigma_k \simeq \frac{e^{-ic_sk\eta}}{\sqrt{2c_sk}}\left(1-\frac{i}{c_sk\eta}\right),
\end{align}
where the time variation of $K$ is assumed to be negligible compared with $a$.
The power spectrum of the original fields on super-horizon scales
are given by
\begin{align}
\mcP_{\dphi} \simeq \frac{H^2}{4\pi^2}, \qquad
\mcP_{\dsig} \simeq \frac{1}{c_s^3 K}\frac{H^2}{4\pi^2}
=\frac{1}{c_s P_X}\frac{H^2}{4\pi^2}.
\label{sigma power spectrum}
\end{align}
Note that the power spectrum of the $\sigma$ field is amplified by the factor of
$(c_s P_X)^{-1}$.

\section{Induced curvature and graviton perturbations}
\label{Induced curvature and graviton perturbations}

In this section, we calculate the  curvature perturbations
and gravitational waves induced by the $\sigma$ field through the third-order terms in the perturbed action.
The third-order action contains many terms, 
\begin{align}
S^{(3)} \supset \int \dd t\dd^3 x a^3 & \Big[
\frac{1}{2} P_X  h_{ij} a^{-2} \partial_i \dsig \partial_j \dsig
-\frac{1}{2} \left(P_{X} -P_{XX} \sig0^2 \right) a^{-2}(\partial_i \dsig)^2 \dN
\notag\\ 
&- \left(\frac{1}{2} P_X  -2P_{XX} \sig0^2\right)(\dsigd)^2\dN
+\frac{1}{2}P_{\sigma\sigma}(\dsig)^2\dN
+\cdots  \Big],
\label{3rd order action}
\end{align}
where we have shown only a few terms. Remember $\dN$ can be written by $\dphi$ and $\dsig$ using eq.~\eqref{dN solution}. 
Actually, there is only one $h(\dsig)^2$ coupling term (the first term in eq.~\eqref{3rd order action}), except for the slow-roll suppressed terms in the gravity sector . However,
there are many $\dphi(\dsig)^2$ coupling terms and it is not transparent which one is  most significant. Then we focus on the term with $\dN(\partial_i \dsig)^2$ (the second term in eq.~\eqref{3rd order action}) because it has a similar form to the graviton coupling term and it is easy to compare them.%
As we see later, the curvature perturbation induced only by this term excludes
the dominant production of gravitational waves via the spectator field.
Thus this treatment is conservative and sufficient.

Since $\sigma$ is a spectator field, 
the comoving curvature perturbation is determined  by the inflaton as%
\begin{equation}
\mcR \simeq  -\frac{H}{\dot{\phi}_0} \delta \phi \simeq
-\frac{2\Mpl H^2}{\dot{\phi}_0^2}\dN \simeq -\frac{\dN}{\epsilon_H}
\label{relation btw R and dN}.
\end{equation}
As we see later, to produce the induced gravitons significantly, $c_s$ should be much smaller than unity. Thus one can approximate
\begin{equation}
c_s^2 = \frac{P_X}{P_X+P_{XX}\sig0^2} \ll 1
\quad\Longrightarrow\quad 
K\equiv P_X+P_{XX}\sig0^2 \simeq P_{XX}\sig0^2 . 
\end{equation}
Then the first line in eq.~\eqref{3rd order action} reads

\begin{equation}
S^{(3)}_{\rm calc} = \int \dd \eta \dd^3 x\, a^2  \Big[\,
\frac{1}{2} P_X  h_{ij}  \partial_i \dsig \partial_j \dsig 
- \frac{1}{2}\epsilon_H K \mcR \partial_i \dsig \partial_i \dsig \Big].
\label{calc action R}
\end{equation}
On the other hand, substituting eq.~\eqref{relation btw R and dN}
into eq.~\eqref{2nd order phi sigma action}, we obtain the relevant second order action as
\begin{align}
S_{\mcR,h}^{(2)} = \int \dd\eta\dd^3 x \left[ a^2\epsilon \Mpl^2
\left( \mcR'^2 - (\partial_i \mcR)^2 \right)
+\frac{a^{2}\Mpl^2}{8}\left(h_{ij}'h_{ij}'-\partial_k h_{ij} \partial_k h_{ij}\right) \right].
\end{align}
Note that all sub-leading terms are dropped and $h_{ij}$ terms come from eq.~\eqref{gravity sector}.
Combining it with  eq.~\eqref{calc action R}, one obtains the equations of motion as
\begin{align}
\mcR''+2\mcH \mcR'-\partial_i^2\mcR &= -\frac{K}{4\Mpl^2} \partial_i \dsig \partial_i \dsig, 
\label{R EoM}
\\
h_{ij}''+2\mcH h_{ij}'-\partial_k^2 h_{ij} &= \frac{2P_X}{\Mpl^2} \tilde{T}^{lm}_{ij} \partial_l \dsig \partial_m \dsig.
\label{h EoM}
\end{align}
Here $\tilde{T}^{lm}_{ij}$ is the projection tensor into the T.T. component defined by
\begin{equation}
\tilde{T}^{lm}_{ij}(\bm{x}) 
\equiv \int \frac{\dd^3 k}{(2\pi)^3} e^{i\bm{k\cdot x}}
\left[
e^+_{ij}(\bm{k})e^+_{lm}(\bm{k})+
e^-_{ij}(\bm{k})e^-_{lm}(\bm{k})
\right].
\end{equation}
Here $e^\pm_{ij}$ are the polarization tensors which are 
written in terms of
 the polarization vectors $e_i(\bm{k})$ and $\bar{e}_i(\bm{k})$ as
\begin{align}
e^+_{ij}(\bm{k}) = \frac{1}{\sqrt{2}}\left[
e_i(\bm{k})e_j(\bm{k})-\bar{e}_i(\bm{k})\bar{e}_j(\bm{k})\right],
\quad
e^-_{ij}(\bm{k}) = \frac{1}{\sqrt{2}}\left[
e_i(\bm{k})\bar{e}_j(\bm{k})+\bar{e}_i(\bm{k})e_j(\bm{k})\right],
\label{linear pol}
\end{align}
where $e_i(\bm{k})$ and $\bar{e}_i(\bm{k})$ are two basis vectors which are orthogonal to each other and $\bm{k}$. 
The only differences between the source terms of $\mcR$ and $h_{ij}$ are
the  coefficients and the projection tensor.
In what follows, we focus on the calculation of $h_{ij}$.
One can solve the equation of $\mcR$ in a similar manner.

Equation~\eqref{h EoM} can be solved 
by the Green function method.
The Green function $g_k(\eta,\tau)$ which satisfies 
\begin{equation}
g_k'' + 2\mcH g_k' +k^2 g_k  = \delta(\eta-\tau),
\end{equation}
is given by
\begin{equation}
g_k (\eta,\tau) = \frac{\theta(\eta-\tau)}{k^3 \tau^2}
\Re e \left[ e^{ik (\eta-\tau)} (1-ik\eta)(-i+k\tau) \right],
\label{Green function}
\end{equation}
where $\theta(\eta)$ is the step function and $\Re e [\cdots]$ represents the real part of $[\cdots]$.
Using this Green function, one finds the inhomogeneous solutions of eq.~\eqref{h EoM} as
\begin{align}
h^\pm_{\bm{k}} (\eta) &= 
\frac{2P_X}{\Mpl^2} 
\int \frac{\dd^3 p\dd^3 q}{(2\pi)^3} \delta(\bm{p+q-k})
e^\pm_{ij}(\bm{k}) p_i q_j
\int^\infty_{-\infty} d\tau
g_k (\eta, \tau)
\sigma_{\bm{p}}(\tau) \sigma_{\bm{q}}(\tau).
\label{Green solution}
\end{align}
Substituting them into the definitions of the power spectrum,
\begin{equation}
\left\langle h^\pm_{\bm{k}}(\eta) h^\pm_{\bm{k'}}(\eta) \right\rangle
= \frac{2\pi^2}{k^3} \delta(\bm{k}+\bm{k'})
\mcP_h^\pm (k,\eta),
\label{power spectrum def}
\end{equation}
one obtains
\begin{align}
\frac{2\pi^2}{k^3} &\delta(\bm{k}+\bm{k'}) \mcP_h^\pm (k,\eta)
=\notag\\
& \frac{4P_X^2}{\Mpl^4} \int \frac{\dd^3 p\dd^3 q\dd^3 p'\dd^3 q'}{(2\pi)^6}
\delta(\bm{p+q-k})\delta(\bm{p'+q'-k'})
e^\pm_{ij}(\bm{k})e^\pm_{ml}(\bm{k'}) p_i q_j p'_m q'_l
\notag \\
&\times\int^\infty_{-\infty} \dd\tau \dd\tau' g_k (\eta,\tau) g_{k'} (\eta,\tau')
\left\langle \sigma_{\bm{p}} (\tau) \sigma_{\bm{q}} (\tau) \sigma_{\bm{p'}} (\tau') \sigma_{\bm{q'}} (\tau') \right\rangle.
\label{4point h}
\end{align}

Since we are treating the source terms of ${\cal R}$ and $h_{ij}$
due to the spectator field $\sigma$ in eqs.~(\ref{R EoM}) and (\ref{h EoM})
as classical stochastic quantities, momentum integrations in eqs.~\eqref{Green solution} and (\ref{4point h}) are performed only in the domain where
the quantum operator $\sigma_{\bm{p}}$ behaves as a classical stochastic
variable.  Specifically we introduce a parameter $\gamma$ smaller than
unity such that  one can approximate
\begin{equation}
\sigma_{\bm{p}}(\eta)\cong \frac{H}{\sqrt{2c_sP_{X}}p^{3/2}} 
 \left( \hat{a}_{\bm{p}} + \hat{a}^\dagger_{\bm{-p}} \right),
\end{equation}
for $|c_s p \eta|<\gamma$, 
where $\hat{a}_{\bm{k}}$ and $\hat{a}^\dagger_{\bm{k}}$ are creation
and annihilation
operators which satisfy the usual commutation relation,
$\left[\hat{a}_{\bm{k}},\hat{a}^\dagger_{\bm{-k'}}\right]
=(2\pi)^3 \delta (\bm{k}+\bm{k'})$.
Then both $\sigma_{\bm{p}}(\eta)$  and
its canonically conjugate momentum variable have the same operator
dependence proportional to $\hat{a}_{\bm{p}} +
\hat{a}^\dagger_{\bm{-p}}$
and commute with each other.

Thus replacing $\sigma_{\bm{p}}$ in eq.~(\ref{4point h}) by
\begin{equation}
\sigma_{\bm{p}}(\eta)\cong \frac{H}{\sqrt{2c_sP_{X}}p^{3/2}} \theta(\gamma+c_s p \eta) \left( \hat{a}_{\bm{p}} + \hat{a}^\dagger_{\bm{-p}} \right),
\end{equation}
one finds
\begin{multline}
\left\langle \sigma_{\bm{p}} (\tau) \sigma_{\bm{q}} (\tau) \sigma_{\bm{p'}} (\tau') \sigma_{\bm{q'}} (\tau') \right\rangle 
\\
=
\frac{H^4}{4P_X^2 c_s^2} (pqp'q')^{-\frac{3}{2}}
\theta(\gamma+c_s p \tau)\theta(\gamma+c_s q \tau)\theta(\gamma+c_s p '\tau')\theta(\gamma+c_s q \tau')
\\ \times
(2\pi)^6 \left[
\delta(\bm{p}+\bm{q'})\delta(\bm{q}+\bm{p'})+\delta(\bm{p}+\bm{p'})\delta(\bm{q}+\bm{q'})
\right].
\end{multline}
Substituting it into  eq.~\eqref{4point h}, and using the symmetry $\bm{p'} \leftrightarrow \bm{q'}$, we obtain
\begin{align}
\mcP_h^\pm (\eta,k) =
\pm&\frac{k^3}{\pi^2 c_s^2} \frac{H^4}{\Mpl^4}
\int \dd^3 p \dd^3 p' \delta(\bm{p-p'-k})
e^\pm_{ij}(\bm{k})e^\pm_{ml}(\bm{k}) \frac{p_i p'_j p'_m p_l}{(pp')^3}
\notag\\
&\times\left[
\int^\infty_{-\infty} \dd \tau g_k (\eta,\tau) \theta(\gamma+c_s p \tau)\theta(\gamma+c_s p' \tau) \right]^2,
\end{align}
where the property of the linear tensor polarization, $e^\pm_{ij}(-\bm{k})=\pm e^\pm_{ij}(\bm{k})$, is used.
The time integration can be analytically performed as
\begin{align}
k^{2}\int^\infty_{-\infty} \dd &\tau g_k (\eta,\tau) \theta(\gamma+c_s p \tau)\theta(\gamma+c_s p' \tau)
\notag \\
&=  1+ \frac{\sin [k(\eta-\tilde{\eta}_{p})]}{k\tilde{\eta}_{p}}
-\frac{\eta}{\tilde{\eta}_{p}} \cos [k(\eta-\tilde{\eta}_{p})],
\label{time integral}
\end{align}
with
\begin{equation}
\tilde{\eta}_{p} \equiv -\frac{\gamma}{c_s \max[p,p']},
\end{equation}
which is the sound horizon crossing time of either $p$ or $p'$ mode,
whichever exits the horizon later.
Finally, in the $p$ integration, one can show that the contribution from
$p\sim \gamma k /c_s$ is dominant.
Then eq.~\eqref{time integral} can be approximated by $1-x \sin(x^{-1})$ for $c_s \ll \gamma$, where $x\equiv c_s p /\gamma k$.
After the angular integration, one finds
\begin{align}
\mcP_h^\pm (\eta,k) 
&\simeq
\pm\frac{8\gamma}{15\pi c_s^3}\frac{H^4}{\Mpl^4} \int_{\epsilon}^{\xi}\dd x
\left[1- x\sin \left(x^{-1}\right) \right]^2,
\label{last integral}
\end{align}
where $\xi \gg 1$ and $\epsilon \ll 1$ are introduced to define the integration interval. Although the $x$ integration cannot
be performed analytically, a  numerical calculation shows that it converges to $\approx1/2$ for a sufficiently large $\xi$ and small $\epsilon$. Remembering $\mcP_h= \mcP_h^+ - \mcP^-_h ,$
one obtains
\begin{equation}
\mcP_h^{(\sigma)} (\eta,k) 
\simeq
\frac{8\gamma}{15\pi c_s^3}\frac{H^4}{\Mpl^4},
\label{final power spectra}
\end{equation}
where the superscript $``(\sigma)"$ denotes that this $\mcP_h$ is 
induced by the $\sigma$ field.
In the same way as $\mcP_h^{(\sigma)}$, one can show the induced power spectrum
of the curvature perturbation is given by
\begin{equation}
\mcP_\mcR^{(\sigma)} (\eta,k) 
\simeq
\frac{\gamma}{32\pi c_s^7}\frac{H^4}{\Mpl^4}.
\label{final R power spectra}
\end{equation}
Thus a spectator field which induces the gravitational waves of eq.~\eqref{final power spectra} inevitably produces the curvature perturbation of eq.~\eqref{final R power spectra} as well.

Since $H\ll \Mpl$, the induced $\mcP_h$, eq.~\eqref{final power spectra},
is negligible compared
to the one coming from the vacuum fluctuation, eq.~\eqref{vac},
unless the  sound speed $c_s$ takes a tiny value 
satisfying $c_s^{3} < 4\pi\gamma H^2/15 \Mpl^2$.
In that case, however, the tensor-to-scalar ratio induced by 
the spectator field,
\begin{equation}
r_\sigma \equiv \frac{\mcP_h^{(\sigma)}}{\mcP_\mcR^{(\sigma)}}
\simeq \frac{256}{15} c_s^4,
\end{equation}
becomes very small.
As a result, the requirement that the induced curvature perturbation
must not exceed the observed value, $\mcP_\mcR^{(\sigma)}\le \mcP_\mcR^{\rm obs}\approx 2.2 \times 10^{-9}$, puts a lower bound on $c_s$ and consequently  constrains $\mcP_h^{(\sigma)}$ as
\begin{equation}
\frac{\mcP_h^{(\sigma)}}{\mcP_h^{\rm vac}}
\, \le\, 2\times 10^{-5} \ \gamma^{\frac{4}{7}}\left(\frac{H}{10^{14}\GeV}\right)^{\frac{2}{7}}.
\end{equation}
As mentioned above, we expect $\gamma\lesssim1$ and it is known $H\lesssim 10^{14}\GeV$ from the CMB observations~\cite{Ade:2013uln, Ade:2013ydc, Ade:2014xna}.
Therefore the induced GW cannot dominate the GW from the vacuum fluctuation.

\section{Interpretation}
\label{interpretation}

In the previous section, it was shown that the spectator field with a tiny
sound speed produces the curvature perturbation $\mcP_\mcR^{(\sigma)}\propto c_s^{-7}$ larger than the gravitational waves, $\mcP_h^{(\sigma)}\propto c_s^{-3}$.
This contrast originates in the difference of coupling constants.
One can see in the right hand side of eqs.~\eqref{R EoM} and 
\eqref{h EoM} that the ratio of the coupling constants is given by
\begin{equation}
\left|\frac{h\dsig^2\ {\rm coupling}}{\mcR\dsig^2\ {\rm coupling}}\right|\simeq \frac{8P_X}{K} =8 c_s^{2}.
\label{coupling ratio}
\end{equation}
%
Hence the $h(\dsig^2)$ coupling is highly suppressed compared to the $\mcR(\dsig)^2$ coupling for $c_s\ll 1$. Now let us take a closer look at what makes these two couplings so different. 

The difference stems from the 
perturbative expansion of the action, $P(X, \sigma) = P+ P_X \delta X +\frac{1}{2} P_{XX} (\delta X)^2 +\cdots$.
The $h(\dsig)^2$ coupling appears  in the perturbation of  $X$, (see eq.~\eqref{X expansion})
\begin{equation}
\delta X \supset \frac{1}{2} a^{-2} h_{ij} \partial_i \dsig\partial_j \dsig.
\end{equation}
Since this is already the third order, no other perturbation can be multiplied to this term. Thus only $P_X\delta X$ carries the $h\dsig^2$ coupling. On the other hand, $\delta X$ also has the following terms:
\begin{equation}
\delta X \supset \sig0^2 \dN  -\frac{1}{2}a^{-2} (\partial_i \dsig)^2,
\end{equation}
where the first term in the right hand side
  is the first order of perturbation and includes $\delta \phi$ (or $\mcR$), while the second term is the second order.
This time, $P_{XX} (\delta X)^2$ can carry the $\mcR (\dsig)^2$ coupling terms. Therefore although the coefficient of the $h(\dsig)^2$ coupling is only $P_X$, the $\mcR (\dsig)^2$ coupling has $P_{XX}\sig0^2$ in its coefficient.
Meanwhile, since the sound speed is given by
\begin{equation}
c_s^2  = \frac{P_X}{P_X+P_{XX}\sig0^2},
\end{equation}
$P_{XX}\sig0^2 \gg P_X$ is necessary to make $c_s$ tiny to boost $\mcP_h^{(\sigma)}$.
However, it results in suppression of
 the $P_{X}$ terms in comparison to the $P_{XX}\sig0^2$ terms. Thus the $h(\dsig)^2$ coupling is suppressed compared to the $\mcR(\dsig)^2$ coupling.

This feature can also be understood as follows. A small sound speed means
that the coefficient of the spacial kinetic term is smaller than that of the time kinetic term. Nevertheless, gravitational wave is induced by the spacial kinetic term of the $\sigma$ field since 
the quadrupole component in the energy momentum tensor of a scalar field
is given only by its spacial kinetic energy.
On the other hand, the adiabatic perturbation is sensitive to both
the time and spacial kinetic energy. Therefore the suppression of the GW
production in comparison with  the curvature perturbation
is a generic consequence of a small sound speed of a scalar field.

In summary, if the sound speed of the spectator field is much smaller than unity, its perturbation is amplified. As a result, both the gravitational waves and the curvature perturbation induced by its second order perturbation are  boosted. However, the sound speed also controls the coupling constants
of the $h(\dsig)^2$ and $\mcR(\dsig)^2$ coupling terms (see eq.~\eqref{coupling ratio}). As $c_s$ becomes smaller, the $h(\dsig)^2$ coupling is more suppressed compared to the $\mcR(\dsig)^2$ coupling. Therefore, since a spectator field with a small sound speed induces the curvature perturbation much more 
than the gravitational waves, it cannot produce the dominant GW
in a way that is consistent with the CMB observation.

\section{Extension to the Galileon theory}
\label{Galileon}

So far the Lagrangian of the spectator field has been assumed to 
take a function of $\sigma$ and $X$ only.
In this section, we show
 that the result obtained in the previous section does not change even if
the action is extended to a more general form.
Specifically we consider the Galileon-like theory~\cite{Nicolis:2008in, Deffayet:2009wt, Deffayet:2010qz, Kobayashi:2010cm},
\begin{equation}
\mathcal{L}_\sigma = P(X,\sigma)-G(X,\sigma)\Box\sigma,
\end{equation}
where $G(X,\sigma)$ is an arbitrary function of $X$ and $\sigma$
and the other pert of action is same as eq.~\eqref{matter lagrangian}.
With this action, the sound speed of $\dsig$ is given by~\cite{Wang:2011dt}
\begin{equation}
c_s^2 =
\frac{P_X+2G_X(\ddot{\sigma}_0+2H\sig0)-2G_\sigma+G_{X\sigma}\sig0^2+G_{XX}\sig0^2\ddot{\sigma}_0}{P_X+6HG_X \sig0-2G_\sigma-G_{X\sigma}\sig0^2+P_{XX}\sig0^2+3HG_{XX}\sig0^3}.
\end{equation}
To make the   sound speed small,  terms proportional to
 $P_{XX}$ or $G_{XX}$ in the denominator have to be much larger 
than the other terms. As we discuss in the previous section, however, that leads to the suppression of the $h(\dsig)^2$ coupling
compared to the $\mcR(\dsig)^2$ coupling because the $h(\dsig)^2$ coupling does
not include $P_{XX}$ nor $G_{XX}$ in the coupling constant while the $\mcR(\dsig)^2$ coupling does.
Indeed, the Galileon term additionally carries the following coupling terms:
\begin{equation}
-N G(X,\sigma)\Box \sigma
\supset
a^{-2}\left(G_\sigma-\frac{3}{2}H\sig0 G_X\right)
h_{ij}\partial_i \dsig \partial_j \dsig
+\frac{3}{2}a^{-2} G_{XX} H\sig0^3 \dN (\partial_i\dsig)^2,
\end{equation}
where we show only the leading terms. It is obvious that the discussion in
Sec.~\ref{interpretation} holds even in this Galileon case.
Therefore we conclude that a spectator field with a small sound speed
cannot produce the dominant GW even if its action is extended to the Galileon theory.
This result implies that it is impossible for {\it a single scalar field with a small sound speed}
to generate GW which is larger than the GW comes from the vacuum fluctuation.

\section{Conclusion}
\label{Conclusion}

It is important to explore an alternative source of primordial GWs
other than GWs from the vacuum fluctuation because it can contribute
the observed GWs and change the consequence on the inflation mechanism.
We consider a spectator scalar field with the a generalized kinetic function
and/or the Galileon-like action which gives it a small sound speed.
The scalar field induces both GWs and curvature perturbation which
are analytically obtained as eq.~\eqref{final power spectra} and 
eq.~\eqref{final R power spectra}, respectively.
Since a small sound speed makes the $\mcR (\delta \sigma)^2$ coupling
much stronger than the  $h (\delta \sigma)^2$ coupling, the induced
curvature perturbation is considerably larger than the induced GWs.
Then the CMB observation put the lower limit on the sound speed, and the stringent constraint on the induced GWs is derived.
Consequently, we conclude that the GWs induced by the spectator scalar
field cannot exceed $\mcP_h^{\rm vac}$.


\section*{Note added}

In the final stage of writing this manuscript a new paper
by Biagetti et al~\cite{Biagetti:2014asa} discussing the same topic
showed up in the arXiv. Their conclusion is basically consistent
with ours. 



\acknowledgments

We would like to thank Takahiro Hayashinaka for useful comments.
This work was supported by the World Premier International
Research Center Initiative (WPI Initiative), MEXT, Japan. 
We acknowledge the support by Grant-in-Aid for JSPS Fellows
No.248160 [T.F.], No. 242775 [S.Y.] and for Scientific Research (B) No. 23340058 [J.Y.].

\end{document}